\documentclass[aps,prl,twocolumn]{revtex4}
\usepackage{epsfig,psfrag}

\begin{document}

\title{Precise Quark Mass Dependence of Instanton Determinant}
\author{Gerald V. Dunne}
\affiliation{Department of Physics, University of Connecticut, Storrs, CT 06269, USA}
\author{Jin Hur}
\author{Choonkyu Lee}
\affiliation{Department of Physics and Center for Theoretical
Physics\\ Seoul National University, Seoul 151-742, Korea}
\author{Hyunsoo Min}
\affiliation{Department of Physics, University of Seoul, Seoul 130-743, Korea\\ Department of Physics, University of Connecticut, Storrs, CT 06269, USA}

\begin{abstract}
The fermion determinant in an instanton background for a quark field of arbitrary mass is determined exactly using an efficient numerical method to evaluate the determinant of a partial wave radial differential operator. The bare sum over partial waves is divergent but can be renormalized in the minimal subtraction scheme using the result of WKB analysis of the large partial wave contribution. Previously, only a few leading terms in the extreme small and large mass limits were known for the corresponding effective action. Our approach works for any quark mass and interpolates smoothly between the analytically known small and large mass expansions.
\end{abstract}
\maketitle

To study instanton-related physics in QCD it is of fundamental importance
to determine the one-loop tunnelling amplitude given by the
Euclidean one-loop effective action in an instanton background \cite{shifman}.
This quantity involves the fermion determinant in a nontrivial gauge configuration and so is also of interest for unquenching effects in dynamical quark simulations in lattice QCD \cite{montvay}.
In the 1970s, 't Hooft \cite{thooft} calculated analytically this one-loop amplitude for {\it massless} scalar or quark fields, but this exact calculation
is no longer possible if the fields have nonzero mass. The small mass limit was extended further by Carlitz and Creamer \cite{carlitz}, and by Kwon et al. \cite{kwon}. This small mass limit is also closely related to the study of zero modes and spectra of the Dirac operator, which have been investigated extensively recently in lattice QCD \cite{bruckmann}. In the other extreme, the large mass
limit is naturally obtained from the Schwinger-DeWitt (or the heat-kernel) expansion
\cite{dewitt,lee,nsvz,fliegner}  within the proper-time representation of
the effective action. For phenomenological applications \cite{shifman}, and also for the extrapolation of lattice results \cite{lattice}
obtained at unphysically large quark masses to lower physical
masses, it is important to be able to connect the large and small mass regimes.
In this paper, we provide the first precise bridge between these two
extremes of small and large mass, by presenting a method which
computes the exact effective action in an instanton background for {\it any} value of the quark mass.
The resulting mass dependence interpolates smoothly between the explicitly known results at the opposite ends.

Our computational procedure is based on an efficient way to compute one-dimensional determinants, combined with the fact that an instanton background is sufficiently symmetric that the corresponding effective action can be reduced to a sum over partial waves (see details below). This bare sum over partial waves is of course divergent, and the nontrivial part of the calculation is to renormalize this divergent sum in a physically unambiguous manner. We use the minimal subtraction scheme  \cite{thooft,kwon} and a WKB analysis \cite{wkb}, which gives the exact counterterms, including finite parts. The result is a simple finite expression for the renormalized one-loop effective action [see (\ref{answer}) below] which can be evaluated numerically.

Due to a hidden supersymmetry  \cite{thooft,jackiw}, the spinor
Dirac operator in an instanton background has the same spectrum
(except for zero modes and an overall multiplicity factor of $4$) as
that of the scalar Klein-Gordon operator in the same instanton
background. This implies that  the one-loop effective action of a
Dirac spinor field of mass $m$ (and isospin $\frac{1}{2}$),
$\Gamma^{F}_{\rm ren}(A;m)$,  is directly related to the
corresponding scalar effective action (for a complex scalar of mass
$m$  and isospin $\frac{1}{2}$) by  \cite{thooft,brown,kwon}:
$\Gamma^{F}_{\rm ren}(A;m)=
-\frac{1}{2}\ln\left(\frac{m^2}{\mu^2}\right) -2\, \Gamma^{S}_{\rm
ren}(A;m)$, where $\mu$ is a renormalization scale. The logarithmic contribution
corresponds to the existence of a zero eigenvalue in the spectrum of
the Dirac operator for a single instanton background. This
relationship (which is special to a self-dual background) has the
important consequence that it is sufficient to consider the scalar
effective action $\Gamma^{S}_{\rm ren}(A;m)$ 
 to learn also about the
corresponding fermion effective action $\Gamma^{F}_{\rm ren}(A;m)$, 
for any mass $m$.

\begin{figure}[h]
\includegraphics[scale=.9]{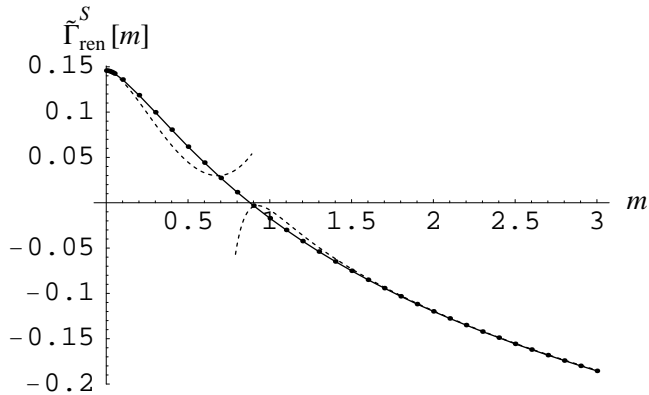}
\caption{Plot of our numercial results for $\tilde{\Gamma}^{S}_{\rm ren}(m)$ from (\protect{\ref{answer}}), compared  with the analytic extreme small and large mass limits [dashed curves] from (\protect{\ref{masslimit}}). The dots are numerical data points from (\protect{\ref{answer}}), and the solid line is a fit through these points.
\label{figure1}}
\end{figure}

As in \cite{thooft}, we consider an SU(2) single instanton of scale $\rho$ :
$A_{\mu}(x) = \frac{\eta_{\mu\nu
a}\tau^{a}x_{\nu}}{r^2+\rho^2}$.
Rather than the scalar effective action $\Gamma^{S}_{\rm ren}(A;m)$, 
it is convenient to consider $\tilde{\Gamma}^{S}_{\rm ren}(m\rho)$, 
a function of  $m\rho$ only,
defined by
\begin{equation}
\Gamma^{S}_{\rm ren}(A;m)=
\frac{1}{6}\ln(\mu\rho)+\tilde{\Gamma}^{S}_{\rm ren}(m\rho) . 
\label{modaction}
\end{equation}
The factor $\frac{1}{6}$ in (\ref{modaction}) is determined by the
one-loop $\beta$ function. We can then concentrate on the $m\rho$
dependence of $\tilde{\Gamma}^{S}_{\rm ren}(m\rho)$, 
and we can set the instanton scale $\rho=1$. In the small and large
mass limits, it is known that $\tilde{\Gamma}^{S}_{\rm ren}(m)$ 
behaves as \cite{nsvz,kwon}
\begin{eqnarray}
\tilde{\Gamma}^{S}_{\rm ren}(m)= &&\label{masslimit}\\
&&\hskip -2cm  
\begin{cases}
{\alpha\left(\frac{1}{2}\right)+\frac{1}{2}\left(\ln m+\gamma-\ln
2\right)m^2 +\dots \quad , \quad m\rightarrow 0 \cr
 -\frac{\ln m}{6}-\frac{1}{75 m^2}-\frac{17}{735 m^4}+\frac{232}{2835 m^6}-\frac{7916}{148225 m^8}+\cdots \cr
 \hspace{5.8cm} , \quad m\rightarrow \infty} 
\end{cases}
\nonumber
\end{eqnarray}
where $\alpha(\frac{1}{2})\simeq 0.145873$, and $\gamma\simeq 0.5772\dots$ is Euler's constant.
These extreme small and large mass limits
(\ref{masslimit}) are shown as dashed curves in Figure \ref{figure1}. There is clearly a significant gap preventing
extrapolation between the small and large mass limits, and for mass between $0.5$ and $1$
(in units of $1/\rho$) neither of these approximations is
particularly accurate. In this paper we provide a simple numerical
procedure to determine $\tilde{\Gamma}^{S}_{\rm ren}(m)$,  
and hence the fermion determinant, for {\it any} value of the mass
$m$, not just asymptotically small or large masses. Our result is
\begin{eqnarray}
\tilde{\Gamma}^S_{\rm ren}(m)&=&\left(\sum_{l=0,\frac{1}{2},\dots}^L \hskip -8pt \Gamma_l^S\right) +2 L^2 + 4 L-\left(\frac{1}{6}+\frac{m^2}{2}\right)\ln L  \nonumber\\
&&\hskip -1cm +\left[\frac{127}{72}-\frac{1}{3}\ln 2+\frac{m^2}{2}-m^2 \ln 2+\frac{m^2}{2}\ln m \right],
\label{answer}
\end{eqnarray}
where $L$ is a large integer. The sum over partial-wave contributions $\Gamma_l^S$ is done numerically, as described below, in Eq. (\ref{lj1}). The other terms in
(\ref{answer}) are renormalization terms, computed using minimal
subtraction and WKB. The renormalized effective action
$\tilde{\Gamma}^S_{\rm ren}(m)$ in (\ref{answer}) is finite,
converges for large $L$,  and can be computed for any mass $m$.
Figure 1 shows that our numerical results provide a very precise
interpolation between the extreme small and large mass limits in
(\ref{masslimit}).

To derive (\ref{answer}), note that the regularized one-loop scalar effective action has  the proper-time representation
\begin{eqnarray}
\Gamma_{\Lambda}^{S}(A;m) &=& - \int_{0}^{\infty}
\frac{ds}{s}(e^{-m^2 s}-e^{-\Lambda^2 s}) \nonumber\\
&&\hskip 1cm \times \int d^4
x\;\textrm{tr}\langle x|{e^{-s(-{\rm D}^2
)}-e^{-s(-\partial^2 )}}|x\rangle \nonumber \\
&\equiv& - \int_{0}^{\infty} \frac{ds}{s}(e^{-m^2 s}-e^{-\Lambda^2
s}) F(s),
 \label{ptaction}
\end{eqnarray}
From this one obtains the
renormalized effective action, in the minimal subtraction scheme \cite{thooft,kwon}:
\begin{eqnarray}
\Gamma^{S}_{\rm ren}(A;m)
&=&
\lim_{\Lambda\rightarrow\infty}
\left[\Gamma_{\Lambda}^{S}(A;m)-\frac{1}{12}
\ln\left(\frac{\Lambda^2}{\mu^2}\right)\right].
\label{renaction}
\end{eqnarray}
Recall that the massive Klein-Gordon (KG) operator, $-D^2+m^2$, for scalars of isospin $\frac{1}{2}$ in the instanton background (with scale $\rho=1$) reduces to the radial form \cite{thooft}
\begin{eqnarray}
{\mathcal M}_{(l,j)} &=&- \frac{\partial^2}{\partial
r^2}-\frac{3}{r}\frac{\partial}{\partial
r}+\frac{4l(l+1)}{r^2}+\frac{4(j-l)(j+l+1)}{r^2+1}\nonumber\\
&&\hskip 1cm  -\frac{3}{(r^2+1)^2} +m^2,
 \label{instop}
\end{eqnarray}
where $l=0,\,
\frac{1}{2},\,1,\,\frac{3}{2},\dots\;$; $j=| l \pm \frac{1}{2}|$, and with degeneracy factor $d_{(l,j)}=(2l+1)(2j+1)$. Without the instanton
background,  the free Klein-Gordon operator is
\begin{equation}
{\mathcal M}^{{\rm free}}_{(l)}=- \frac{\partial^2}{\partial
r^2}-\frac{3}{r}\frac{\partial}{\partial r}+\frac{4l(l+1)}{r^2} +m^2,
\label{freeop}
\end{equation}
with degeneracy factor $d_{(l)}=(2l+1)^2$. 
Thus,
it is natural \cite{wkb} to combine the partial waves $(l,j=l+\frac{1}{2})$ and $(l+\frac{1}{2},j=l)$, which have common degeneracy factor $(2l+1)(2l+2)$, so that the sum over $l$ and $j$ reduces to a sum over $l$. This sum is, however, divergent and we must define a consistent regularization and renormalization. To this end, we use the proper-time regularization in (\ref{ptaction}), and split the $l$ sum as follows:
\begin{eqnarray}
\Gamma_\Lambda^S(A; m)&=& \hskip -5pt \sum_{l=0,\frac{1}{2},\dots}^L \hskip -5pt
\Gamma_{\Lambda, (l)}^S(A; m) + \hskip -5pt \sum_{l=L+\frac{1}{2}}^\infty \hskip -5pt \Gamma^S_{\Lambda, (l)}(A; m) 
\label{actionsplit}
\end{eqnarray}
where $L$ is a large but finite integer. In the first sum, which is finite, the cutoff $\Lambda$
is irrelevant and we can use a numerical method [described below] to
evaluate the sum. For the second sum, we use a combination of the
WKB approximation, which is good for large $l$, and Euler-Maclaurin
summation to perform the sum.

To begin, we consider the first sum in (\ref{actionsplit}):
\begin{eqnarray}
\sum_{l=0,\frac{1}{2},\dots}^L \Gamma_l^S(A; m) & =& \sum_{l=0,\frac{1}{2}, \dots}^L (2l+1)(2l+2) \nonumber\\
&&
\hskip -3cm \times
\left\{ \ln \det \left(\frac{{\mathcal M}_{(l,l+\frac{1}{2})}}{{\mathcal M}^{{\rm free}}_{(l)}}\right)+  \ln \det \left(\frac{{\mathcal M}_{(l+\frac{1}{2},l)}}{{\mathcal M}^{{\rm free}}_{(l+\frac{1}{2})}}\right)\right\}
\label{lj}
\end{eqnarray}
These one-dimensional determinants can be computed efficiently using the following result \cite{levit,coleman,dreyfus,forman,kirsten}: Suppose ${\mathcal M}_1$ and ${\mathcal M}_2$ are two second order ordinary differential operators on the interval $r\in [0,\infty)$, with Dirichlet boundary conditions. Then the ratio of the determinants of ${\mathcal M}_1$ and ${\mathcal M}_2$ is given by
\begin{eqnarray}
\det\left(\frac{{\mathcal M}_1}{{\mathcal M}_2}\right)=\lim_{R\to\infty}\left(\frac{\psi_1(R)}{\psi_2(R)}\right)
\label{theorem}
\end{eqnarray}
where $\psi_i(r)$ ($i=1,2$) satisfies the initial value problem
\begin{eqnarray}
{\mathcal M}_i\, \psi_i(r)=0\quad ;  \psi_i(r=0)=0\,\, ;  \psi^\prime_i(r=0)=1
\label{ode}
\end{eqnarray}
Since an initial value problem is very simple to solve numerically, this theorem provides an efficient way to compute the determinant of an ordinary differential operator. Note in particular that no direct information about the spectrum (either bound or continuum states, or phase shifts) is required in order to compute the determinant.

We can simplify the numerical
computation further because for the free KG operator, ${\mathcal
M}^{{\rm free}}_{(l)} $, the solution to (\ref{ode}) is just the modified Bessel function: $ \psi^{\rm free}_{(l)}(r)=\frac{I_{2l+1}(m
r)}{r}$. Since this grows exponentially fast at large $r$, this should also
be true of the numerical solutions of (\ref{ode}) for the instanton
operators, ${\mathcal M}_{(l,j)} $, 
in (\ref{instop}). Thus, it is numerically better to consider the
ODE satisfied by the {\it ratio} (a similar idea was used by Baacke
et al in their analysis of metastable vacuum decay \cite{baacke}) :
$T_{(l,j)}(r)=\frac{\psi_{(l,j)}(r)}{\psi^{\rm free}_{(l)}(r)}$.
In fact, since we are ultimately interested in the logarithm of the determinant, it is more convenient (and more numerically stable) to consider $S_{(l,j)}=\ln T_{(l,j)}$, which satisfies
\begin{eqnarray}
&&\frac{d^2 S_{(l,j)}}{dr^2}+\left(\frac{d S_{(l,j)}}{dr}\right)^2+\left(\frac{1}{r}+2m\frac{I^\prime_{2l+1}(m r)}{I_{2l+1}(m r)}\right)\frac{d S_{(l,j)}}{dr}\nonumber\\
&&\hskip 1cm = \frac{4(j-l)(j+l+1)}{r^2+1}-\frac{3}{(r^2+1)^2}
\label{nonlinear}
\end{eqnarray}
where $S_{(l,j)}(r)$ satisfies the initial conditions $S_{(l,j)}(0)=0$ and $S^\prime_{(l,j)}(0)=0$.
Thus, the first, finite, sum in (\ref{actionsplit}) for the bare effective action can be evaluated as
\begin{eqnarray}
\sum_{l=0,\frac{1}{2},\dots}^L \Gamma_l^S(A; m)&=&\sum_{l=0,\frac{1}{2}, \dots}^L (2l+1)(2l+2)  \nonumber\\
&&\hskip -2cm \times \left\{ S_{(l,l+\frac{1}{2})}(r=\infty)+S_{(l+\frac{1}{2},l)}(r=\infty) \right\}.
\label{lj1}
\end{eqnarray}
\begin{figure}[h]
\includegraphics[scale=0.5]{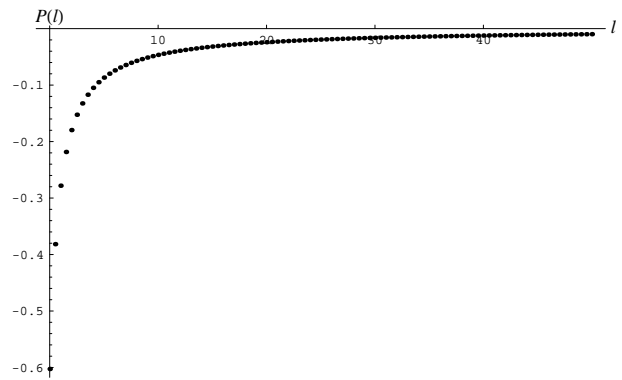}
\caption{Plot of the $l$ dependence of $P(l)\equiv S_{(l,l+\frac{1}{2})}(r=\infty)+S_{(l+\frac{1}{2},l)}(r=\infty)$,  for $m=1$. The sum of these two asymptotic values is $O(\frac{1}{l})$ for large $l$.
\label{figure2}}
\end{figure}
The large $r$ values of $S_{(l,l+\frac{1}{2})}(r)$ and $S_{(l+\frac{1}{2},l)}(r)$ can be extracted with excellent precision.
In fact, the asymptotic values of  $S_{(l,l+\frac{1}{2})}(r)$ and
$S_{(l+\frac{1}{2},l)}(r)$ very nearly cancel one another, and for a
given mass, as a function of $l$, the combination
$[S_{(l,l+\frac{1}{2})}(r=\infty)+S_{(l+\frac{1}{2},l)}(r=\infty)]$
falls off 
in magnitude like $\frac{1}{l}$ -- see Figure \ref{figure2}. Thus,
the sum in (\ref{lj1}) will have terms going like, $L^2$, $L$ and
$\ln L$, as well as terms finite and vanishing for large $L$. We now
show that these potentially divergent terms are exactly canceled by
terms in the second sum in (\ref{actionsplit}).

The second sum in (\ref{actionsplit}) can be analyzed using the Euler-Maclaurin method \cite{bender} as follows.  Write
\begin{eqnarray}
\sum_{l=L+\frac{1}{2}}^\infty \hskip -5pt \Gamma^S_{\Lambda, (l)}(A; m)\hskip -2pt  =\hskip -4pt \int_0^\infty \hskip -8pt ds
\left[-\frac{1}{s}\left(e^{-m^2 s}-e^{-\Lambda^2
s}\right)f(s) \right] \label{second} 
\end{eqnarray}
Using WKB, which is good for large $l$,  we can write \cite{wkb}
\begin{eqnarray}
f(s)=\int_0^\infty dr \left(\sum_{l=L+\frac{1}{2}}^\infty 
f_l(s,r)\right) \label{wkb}
\end{eqnarray}
where for each $l$, $f_l(s,r)$ has a local expansion in terms of the Langer-modified potential \cite{langer}
\begin{eqnarray}
V_{(l,j)}^{\rm Lang}\hskip -2pt =\hskip -2pt \frac{(2l+1)^2}{r^2}+\frac{4(j-l)(j+l+1)}{r^2+1}-\frac{3}{(r^2+1)^2}
\label{langer}
\end{eqnarray}
The first three orders of the  WKB approximation for $f_l$ were computed in \cite{wkb}. We find $f_l(s,r)=(2l+1)(2l+2)[f_{(l,l+\frac{1}{2})}(s,r)+f_{(l+\frac{1}{2},l)}(s,r)]$, where to first order
\begin{eqnarray}
f_{(l,j)}^{(1)}(s,r)=\frac{1}{2\sqrt{\pi s}}\, e^{-[s V_{(l,j)}^{\rm Lang}(r)]},
\label{wkbfirst}
\end{eqnarray}
and to second order
\begin{eqnarray}
f_{(l,j)}^{(2)}(s,r)\hskip -2pt =\hskip -2pt  \frac{1}{2\sqrt{\pi s}}\hskip -2pt\left(\frac{s}{4 r^2}-\frac{s^2}{12}\frac{d^2 V_{(l,j)}^{\rm Lang}}{dr^2}\right)\hskip -3pt  e^{-[s V_{(l,j)}^{\rm Lang}(r)]}
\label{wkbsecond}
\end{eqnarray}
The sum over $l$ in (\ref{wkb}) can now be performed using the Euler-Maclaurin expansion \cite{bender} :
\begin{eqnarray}
\sum_{l=L+\frac{1}{2}}^\infty f_l\, = 2 \int_L^\infty dl\, f(l)-\frac{1}{2} f(L)-\frac{1}{24}f^\prime(L)+\dots
\label{euler}
\end{eqnarray}
For each order of the WKB expansion, the $l$ dependence is such that the $l$ integral in
(\ref{euler}) can be done exactly. Then in the large $L$ limit, the $r$ and $s$ integrals in (\ref{wkb})
and (\ref{second}) can also be done, leading to \cite{inprep}
\begin{eqnarray}
\sum_{l=L+\frac{1}{2}}^\infty \hskip -5pt  \Gamma^S_{\Lambda, (l)}&\sim& \frac{1}{6}\ln \Lambda+2 L^2 + 4 L-\left(\frac{1}{6}+\frac{m^2}{2}\right)\ln L \nonumber\\
&& \hskip -2.5cm +\left[\frac{127}{72}-\frac{1}{3}\ln 2+\frac{m^2}{2}-m^2 \ln
2+\frac{m^2}{2}\ln m \right]+\dots
\label{div}
\end{eqnarray}
Only the first two orders of the WKB expansion are required for this result; all higher orders contribute terms vanishing in the large $L$ limit. It is important to identify the physical role of the various terms in (\ref{div}). The first term is the expected $\frac{1}{6}\ln \Lambda$ term canceled in (\ref{renaction}). The next three terms give the quadratic, linear and logarithmic divergences in $L$ which cancel the corresponding divergences in the first sum in (\ref{actionsplit}), which were found in our numerical data -- see Fig. \ref{figure2}. It is a highly nontrivial check on the WKB computation that these terms have the correct coefficients to cancel these divergences. Combining (\ref{div}) and (\ref{lj1}), we obtain our result (\ref{answer}).

As an interesting application, our result (\ref{answer}) provides a very simple analytic computation of 't Hooft's leading small mass result. When $m=0$, the numerical integration can be done exactly and one finds that \cite{inprep}
\begin{eqnarray}
S_{(l,l+\frac{1}{2})}(r=\infty)+S_{(l+\frac{1}{2},l)}(r=\infty)=\ln\left(\frac{2l+1}{2l+2}\right)
\label{exact}
\end{eqnarray}
Then it follows from (\ref{answer}) that
\begin{eqnarray}
\tilde{\Gamma}^S_{\rm ren}(m=0)&=&\sum_{l=0,\frac{1}{2},\dots}^L (2l+1)(2l+2)\ln\left(\frac{2l+1}{2l+2}\right)
\nonumber\\
&&+2 L^2 + 4 L-\frac{1}{6}\ln L  +\frac{127}{72}-\frac{1}{3}\ln 2 \nonumber\\
&=& -\frac{17}{72}-\frac{1}{6}\ln 2 +\frac{1}{6}-2\zeta^\prime(-1)+O(\frac{1}{L})\nonumber\\
& \buildrel{L\to\infty}\over \longrightarrow & 
\alpha\left(\frac{1}{2}\right)= 0.145873... \label{masslessanswer}
\end{eqnarray}
which agrees precisely with the leading term in the small mass limit
in (\ref{masslimit}).

For other, nonzero, values of the mass $m$ we found excellent convergence in computing $\tilde{\Gamma}^S_{\rm ren}(m)$ in (\ref{answer}) with $L=50$, combined with Richardson extrapolation \cite{bender}. In Figure \ref{figure1} these results are compared to the analytic large and small mass expansions in (\ref{masslimit}) --
the agreement is remarkable. Thus, our expression (\ref{answer}) provides a simple and numerically exact interpolation between the large mass and small mass regimes.

In conclusion, beyond the instanton problem considered in this paper, our approach should provide a practical numerical scheme for the one-loop effective action in a broad class of nonabelian background fields.

\vskip .5cm
{\bf Acknowledgments:} GD thanks T. Blum,
H. Gies, V. Khemani and  K. Kirsten for helpful discussions and correspondence, and the US DOE for support through the grant DE-FG02-92ER40716. The work of CL was supported by the Korea Science Foundation ABRL program (R14-2003-012-01002-0). HM thanks the University of Connecticut for hospitality during sabbatical, and the University of Seoul for support.

\end{document}